\documentclass[aps,pre,superscriptaddress,floatfix,tightenlines,showpacs,twocolumn,notitlepage]{revtex4-1}
\usepackage{eqnarray,amsmath}
\usepackage{xcolor}
\usepackage{graphicx}
\usepackage{mathtools}

\usepackage{amsfonts}	
\usepackage{float}		
\usepackage{longtable}
\usepackage{multirow}
\usepackage{array}
\usepackage{hyperref}
\newcommand{\norm}[1]{\left\vert#1\right\vert}

\graphicspath{{./figures/}}			
\begin{document}
	\preprint{AIP/123-QED}
	\title{Dynamic Scaling in Rotating Turbulence: A Shell Model Study}
	\author{Shailendra K. Rathor}
	\email{skrathor@iitk.ac.in}
	\affiliation{Department of Physics, Indian Institute of Technology Kanpur, Uttar Pradesh 208016, India}
	\author{Sagar Chakraborty}
	\email{sagarc@iitk.ac.in}
	\affiliation{Department of Physics, Indian Institute of Technology Kanpur, Uttar Pradesh 208016, India}
	\author{Samriddhi Sankar Ray}
	\email{samriddhisankarray@gmail.com}
	\affiliation{International Centre for Theoretical Sciences, Tata Institute of Fundamental Research, Bangalore 560089, India}
	\begin{abstract}

We investigate the scaling form of appropriate time-scales extracted from time-dependent correlation functions
		in rotating, turbulent flows. In particular, we obtain 
		precise estimates of the dynamic exponents $z_p$, associated with the 
		time-scales, and their relation with the more commonly measured equal-time 
		exponents $\zeta_p$. These theoretical predictions, obtained by using the multifractal
		formalism, are validated through extensive numerical
		simulations of a shell model for such rotating flows. 
	\end{abstract} 

	\maketitle 

	Many aspects of turbulence are understood through
	$p$-th order correlation functions of velocity increments across
	suitably defined length scales $r$ which lie in the so-called
	\textit{inertial} ranges of the
	flow~\cite{Monin1971,Monin1975,Frisch1995}. In simple terms, the
	inertial range is well-separated from, and lie between, the
	system-dependent energy injection scale $L$ and dissipation scale
	$\eta$ of the turbulent flow. We now know that there exist  
	power-laws~\cite{Kolmogorov1941a,Kolmogorov1941b,Frisch1995} in these
	correlators---typically called \textit{structure functions}---and a
	universality of the associated scaling exponents $\zeta_p$ which are
	perhaps universal for a given class of turbulent flows but may well
	vary for different forms of turbulence. Thus, the evidence favouring
	the universality of such exponents in fully developed, homogeneous and
	isotropic~\cite{Jensen1991,Frisch1995,Sreenivasan_1997},
	passive-scalar~\cite{Kraichnan1994,Jensen1991,Wirth1996,Warhaft_2000},
	magnetohydrodynamic (MHD)~\cite{Verma2004,Sahoo2011,Banerjee2013},
	two-dimensional~\cite{Boffetta2002,Ott2005,Ray2011,Boffetta_2012} and
	indeed rotating
	turbulence~\cite{Mueller2007,Seiwert2008,Thiele2009,Rathor2020}, to
	name a few, is overwhelming; nevertheless, the values of $\zeta_p$ are
	known to be different and specific to each of these turbulent flows. 

The algebraic nature of these structure functions, and indeed the universality
of the exponents, also reminds one of the behaviour of correlation functions
near a critical point~\cite{chaikin}, for example in spin systems. However, for
the turbulent flows that we are familiar with this analogy is
limited~\cite{Pandit-Review}. This is because in fully developed turbulence an
infinite set of exponents are required to fully characterise different-order
structure functions in the inertial range as opposed to the simple scaling one
is familiar with in critical phenomena. Such a complexity, which can be rationalised through a 
multifractal description of turbulent flows, has also meant that unlike in critical
phenomenon where studies of static and dynamic correlators~\cite{Hohenberg1977}
went more or less hand-in-hand, the study of time-dependent structure functions
is more recent in turbulence. 

Nevertheless, over the past decades there has been a concerted effort to
generalise the \textit{dynamic-scaling ansatz} in the critical phenomenon---namely
the dynamic scaling exponent $z$ associated with the relaxation time $\tau$
near a critical point---and obtain estimates of the multiscale nature of
time-dependent structure functions in turbulence. These investigations have, however,
been limited to homogeneous, isotropic turbulence in two and three dimensions
or for the case of passive-scalar turbulence~\cite{Mitra2003,Mitra2004,Mitra2005,Ray2008,Biferale2011,Ray2011}. In particular, these studies
demonstrate that just like the case of equal-time exponents $\zeta_p$, there
exists an infinite set of (universal) exponents $z_p$  whose values depend on the \textit{kind} of relaxation
time fished out from the order-$p$, time-dependent structure functions. 
Perhaps the most important success of these studies was the generalisation of the 
Frisch--Parisi multifractal formalism~\cite{Frisch1985} for the velocity field ${\bf u}$ to derive (linear) 
bridge relations~\cite{Mitra2004,Hayot1998,Lvov1997} connecting the dynamic $z_p$ and equal-time $\zeta_p$ exponents 
and establish the notion of \textit{dynamic multiscaling}.

The complex nature of time correlations in these systems is intrinsic. But what
happens when there is an external \textit{global} time-scale governing the
statistical nature of the turbulent flow itself? Indeed for such systems, it is
difficult to separate the hierarchy of dynamics intrinsic to the system and the
time-correlations set by the imposed time-scale making the study of the 
nature of dynamic (multi)scaling, when such effects are at play, non trivial. 

One of the more natural and ubiquitous examples of turbulence with an imposed
global time-scale is that of rotating
turbulence~\cite{Greenspan1968,Ibbetson1975,davidson_2013,Godeferd2015},
observed in geophysical phenomena ~\cite{Aurnou_2015} including oceanic and
atmospheric flows ~\cite{Cho2008}, astrophysical phenomena ~\cite{Barnes_2001},
and many engineering applications. 
\textcolor{black}{When the Coriolis force dominates over the nonlinear term, 
strongly rotating, but mildly turbulent, three-dimensional flows 
tend to become two-dimensional, consistent with the Taylor--Proudman
theorem~\cite{davidson_2013,Taylor_1917}.} However, when the flow becomes
turbulent, the nonlinear effects can no longer be
ignored~\cite{Metais1996,Smith_1999,Thiele2009,Mininni2009,Yarom_2013,Sharma2018,Sharma2018a};
in fact, the nonlinear interactions among the inertial waves play an important
role in developing the quasi-two-dimensional behaviour of rotating
turbulence~\cite{Hopfinger1982,Cambon1989,Biferale2016,Galtier2003,Cambon_2004,Bewley_2007,Reun_2017}.

In such rotating turbulent flows, the addition of a global rotation rate
$\Omega$ through the Coriolis force sets a unique time scale $1/\Omega$. At the
level of statics, we know~\cite{Zeman1994} that this time-scale leads to a
characteristic length scale in the problem: The Zeman scale $\ell_{\Omega} =
\sqrt{\varepsilon/\Omega^3}$, where $\varepsilon$ is the mean kinetic energy
dissipation. The role of this global rotation, via the Zeman scale, in
determining the equal-time statistics of three-dimensional rotating turbulence
has been extensively studied. In particular, we know that unlike homogeneous
and isotropic turbulence, in the limit of large Reynolds numbers, when $L \gg
\ell_{\Omega} \gg \eta$, the equal-time (longitudinal) structure functions of
the (projected) velocity increments $\delta u (r) = [\bf{u}({\bf x} +
{\bf r}) -  \bf{u}({\bf x})]\cdot {\bf \hat{r}}$ in the inertial range
$L \gg r \gg \eta$ show a dual scaling; 
${\bf \hat{r}}$ is the unit vector along the separation vector ${\bf r}$. 

More precisely, defining the $p$-th order, equal-time structure function
$\mathcal{S}_p(r) = \langle |\delta u(r)|^p\rangle$, the equal-time exponents
are extracted via the power-law $\mathcal{S}_p(r) = r^{\zeta_p}$ in the
inertial range. The rotation-induced Zeman scale results in two different
classes of exponents: Theoretical estimates suggest that for the
rotation-dominated larger scales $L \gg r \gg \ell_{\Omega}$, the exponents
$\zeta_p = p/2$; however, at smaller scales $\ell_\Omega \gg r \gg \eta$, which
are less sensitive to the Coriolis effects, $\zeta_p = p/3$ as is the case for
fully developed three-dimensional
turbulence~\cite{Mueller2007,Seiwert2008,Thiele2009,Rathor2020}.  A consequence
of this is that kinetic energy spectrum $E(k)$ also displays
dual-scaling~\cite{Zeman1994,Zhou1995,Chakraborty2010,Chakraborty_2007}: $E(k)
= k^{-2}$ and $E(k) = k^{-5/3}$ for wavenumbers smaller or larger,
respectively, than the Zeman wavenumber $k_\Omega = 1/\ell_\Omega$. \textcolor{black}{This dual 
scaling of the energy spectra is seen remarkably well in shell models, such as the one we use here, as 
shown in Fig. 1 of Ref.~\cite{Rathor2020}.} Of course,
measurements suggest strong intermittency corrections to this simple
\textit{dimensional} form. Thus, in rotating turbulence---just like in
homogeneous, isotropic turbulence---there exists multiscaling at the level of
equal-time statistics. 

But, is there a similar multiscaling for the dynamic correlators in such
systems?  This remains a somewhat open question because while different aspects of
Lagrangian turbulence of rotating flows have been
studied~\cite{Cambon2004,Castello_2011,Biferale2016,Yarom2017,Maity2019},
studies of dynamic correlators are sparse~\cite{Favier_2010,Leoni2014}. Furthermore,
these studies~\cite{Favier_2010,Leoni2014} use an Eulerian approach to measure the second-order dynamic
correlation function which, as we know from insights developed in non-rotating
turbulence~\cite{Mitra2004}, can lead to an oversimplification and mask an
underlying multiscaling as we illustrate below. 

In the much simpler non-rotating, homogeneous, isotropic turbulent flow, a
na\"ive calculation of dynamic scaling within the Eulerian framework---in a
manner similar to what is done for equal-time structure functions---yields a
trivial dynamic exponent of unity because the sweeping effect dominates and
thus, linearly couples the temporal and spatial scales. Indeed, this
``sweeping'' effect leads to the simpler dynamic exponents for the Eulerian
time-dependent correlation functions in rotating turbulence as well as reported
by Favier, Godeferd and Cambon~\cite{Favier_2010}. 

Thus, unlike for equal-time structure functions, special
care must be taken which eliminates this sweeping effect in order to obtain
non-trivial dynamic (multi)scaling exponents. This can be done through the
Lagrangian or the quasi-Lagrangian
framework~\cite{Lvov1997,Biferale1999,Kaneda1999,Mitra2003,Mitra2004,Mitra2005,Ray2008,Pandit_2008}.
While the former allows us to measure the structure functions of temporal
velocity increments $\delta {\bf u}(\tau) = {\bf u}(t+\tau) - {\bf u}(t)$, the latter 
\textcolor{black}{is especially useful as it allows us to obtain time-dependent
structure functions for velocity increments and hence, in the limiting case, recovers the 
(Eulerian) equal-time structure functions~\cite{Lvov87,Lvov1997} as we now show. 
The quasi-Lagrangian velocity field
\begin{equation}
\textbf{v}(\textbf{r}_0,t_0|\textbf{x},t_0 + t) \equiv \textbf{u}(\textbf{x} +
\textbf{R}_L(\textbf{r}_0,t_0|t_0 + t),t_0 + t)
\end{equation}
is measured along the Lagrangian trajectory  
$\textbf{R}_L(\textbf{r}_0,t_0|t_0 + t)$ of a fluid particle starting at $(\textbf{r}_0, t_0)$. This allows 
us to define the (quasi-Lagrangian) velocity increments $\delta v(r,t)  =
[\textbf{v}(\textbf{r}_0,t_0|\textbf{x} + \textbf{r} ,t_0 + t) -
\textbf{v}(\textbf{r}_0,t_0|\textbf{x},t_0 + t)]\cdot {\bf \hat r}$ and thence the 
time-dependent structure function $\mathcal{F}_p(r,{t_1,...,t_p}) \equiv \langle \delta v (r, t_1) \delta
v (r, t_2) ...\delta v (r, t_p) \rangle$. By setting $t_1 = t_2 =
\cdots= t_p = t$, the quasi-Lagrangian time-dependent structure function is
written simply as $\mathcal{F}_p(r,t)$ with the obvious identity
$\mathcal{F}_p(r,t = 0) \equiv \mathcal{S}_p(r)$.}

The quasi-Lagrangian structure function also lends itself to an adaptation of the
Frisch-Parisi multifractal formalism~\cite{Frisch1985,Frisch1995} for the equal-time structure function.
Assuming a multifractal description of rotating turbulence, the velocity field
ought to possess a range of (universal) scaling exponents $h \in \mathcal{I}
\equiv (h_{\rm min}, h_{\rm max})$, each of which corresponds to a fractal set
$\Sigma_h$ of dimension $\mathcal{D}(h)$.  This allows one to write down the
velocity increments ${\delta u({\bf x}, r)}/{u_L} \propto 
({r}/{L})^h$, where $u_L$ is the velocity associated with the large
length scale of the flow. Given the multifractal description, for individual
increments it is important to keep track of the point ${\bf x}$ at which the
increments are taken because the increment picks up different scaling exponents
$h$ for every ${\bf x} \in \Sigma_h$.

Such a prescription allows us to define the equal-time structure function in terms of the scaling exponents $h$ and the measure $d\mu(h)$ which 
gives the \textit{weight} of the contributing fractal sets:
	\begin{equation}
	\label{eqn:sf}
		S_p(r) \propto u^p_L \int_\mathcal{I} d\mu(h) \left(\frac{r}{L}\right)^{ph + 3 - \mathcal{D}(h)}.
	\end{equation}
Formally, the measured scaling exponents $\zeta_p$ are then extracted through a saddle-point calculation.

We now extend the equal-time formalism for time-dependent structure functions
	\begin{equation}
	\label{eqn:tdsf}
		\mathcal{F}_p(r,t) \propto u^p_L \int_{\mathcal{I}} d\mu(h) \left(\frac{r}{L}\right)^{ph + 3 - \mathcal{D}(h)} \mathcal{G}^{p,h}\left(\frac{t}{\tau_{p,h,\Omega}(r)}\right), 
	\end{equation}
where $\tau_{p,h,\Omega}(r)$ is the characteristic scale-dependent time scale of the flow. 
The scaling function $\mathcal{G}^{p,h}$ is unity at $t = 0$ and its integral is assumed to exist to allow us to define 
the integral-time scale: 
	\begin{equation}
	\label{eqn:int_time}
		\mathcal{T}^I_{p}(r) = \left[ \frac{1}{\mathcal{S}_p(r)} \int_{0}^{\infty} \mathcal{F}_p(r,t) dt \right] \textcolor{black}{\sim r^{z_p}}. 
	\end{equation}

In order to proceed further and calculate the dynamic exponent $z_p$, we make
reasonable assumptions on the time-scale $\tau_{p,h,\Omega}$.
The phenomenology of rotating turbulence suggests that in the
rotation-dominated regime $L \gg r \gg \ell_{\Omega}$, to leading order, the time-scale
is set by the rotation rate $\Omega$ and hence $\tau_{p,h,\Omega} \propto
{1}/{\Omega}$. On the other side of the Zeman scale  $\ell_\Omega \gg r \gg
\eta$ however, we expect $\tau_{p,h,\Omega} \equiv \tau_{p,h} \propto r^{1-h}$
consistent with the ideas of homogeneous and isotropic turbulence. 

By using standard tools to evaluate the integral in Eq.~\eqref{eqn:int_time}, we eventually obtain 
\textcolor{black}{(see, e.g., Refs.~\cite{Lvov1997,Mitra2004})} 
	\begin{equation}
	z_p \sim 
	\left\{
	\begin{array}{ll}
	1 + (\zeta_{p-1} - \zeta_p), &  \ell_\Omega \gg r \gg \eta;\\
	0, & L \gg r \gg \ell_{\Omega}.
	\end{array} 
	\right.
		\label{exponents}	
	\end{equation}
Furthermore, the same analysis suggests that for  $L \gg r \gg \ell_{\Omega}$, the 
integral-time scale $\mathcal{T}^I_{p}(r) \propto 1/\Omega$. \textcolor{black}{Indeed, 
this form is perhaps not entirely surprising given the scale-independent form of  $\tau_{p,h,\Omega}$.}

Our predictions suggest that in rotating turbulence, the time-dependent
structure functions also show dual scaling consistent with what we know from
equal-time measurements. Indeed, the bridge relation connecting the
integral-time scale based dynamic exponent $z_p$ for scales where turbulent
fluctuations swamp the effect of rotation is identical to what happens in
three-dimensional turbulent flows~\cite{Mitra2004}. On the other hand, the
dynamic structure functions are scale-independent ($z_p = 0$) as soon as
rotation is dominant.  (It is perhaps useful to keep in mind that although to
leading order our analysis shows the integral-time scale in the
rotation-dominated regime is scale-independent, the structure
functions themselves are not as we show below.)

Are our results surprising? The surprise and apparent contradiction arises
when we examine the dynamics in terms of local turn-over time-scales of the
flow $\mathcal{T}^{\rm local}(r) \sim r/\delta u(r)$. For scales smaller than 
the Zeman scale by using $\delta u(r) \sim r^{1/3}$, we obtain the local $p$-independent dynamic exponent 
$z^{\rm local} = 2/3$. This exponent is exactly the same as what we obtain from Eq.~\eqref{exponents} in 
the absence of intermittency correction, i.e., $\zeta_p = p/3$. However, 
for scales larger than $\ell_\Omega$, a similar analysis yields $z^{\rm local} = 1/2$ since $E(k) \sim k^{-2}$. This 
result is in sharp contradiction with the exponent $z_p = 0$ as obtained above \eqref{exponents} but also, crucially, suggests 
that in the rotation-dominant regime, the dynamic structure function is scale-dependent. 

This begs the question as to which of these two approaches are correct and how is this contradiction resolved.
Indeed, how valid are our theoretical predictions~\eqref{exponents} when confronted with data from simulations? 

While formally quasi-Lagrangian structures are well-defined, measurements from
direct numerical simulations (DNSs) of the three-dimensional Navier-Stokes
equation are still a challenge~\cite{Biferale2011,Ray2011}. Fortunately, this
problem of circumventing sweeping through a quasi-Lagrangian description was
solved~\cite{Mitra2004} by adopting a shell model approach. Indeed, by
construction, shell models are dynamical systems for (complex) variables which
resemble velocity increments and sweeping is eliminated by restricting the
coupling between modes which are only nearest or next-nearest neighbours.
Remarkably, such a dynamical systems approach to
turbulence~\cite{Frisch1995,Jensen1999,Biferale-review,Jensen1991,BohrBook}
does capture the essential multifractal and cascade processes of fully
developed turbulence as was recognised since the pioneering works of
Obukhov~\cite{Obukhov1974}, Desnyansky and Novikov~\cite{Desnyansky1974},
Gledzer~\cite{Gledzer1973}, and Ohkitani and Yamada~\cite{Ohkitani1982} and then generalised 
to several other single and multiphase flows~\cite{Aurell_1994,Giuliani_2002,Frick1998,Basu1998,Giuliani_1998,Brandenburg_1996,
Hori2005,Galtier_2008,Banerjee2013,Wirth1996,Mitra2005,Kalelkar2005,Ray2011,
Wacks_2011,Boue2012,Boue2013,Shukla_2016}. 
Moreover, shell
models, although structurally isotropic, reproduce and predict
many properties of the rotating turbulence, e.g., two-dimensionalisation, the
dual scaling of energy spectrum, and the scaling of equal-time structure
functions ~\cite{Hattori2004,Chakraborty2010,Rathor2020}.

	\begin{figure*}
		\includegraphics[width=1.0\linewidth]{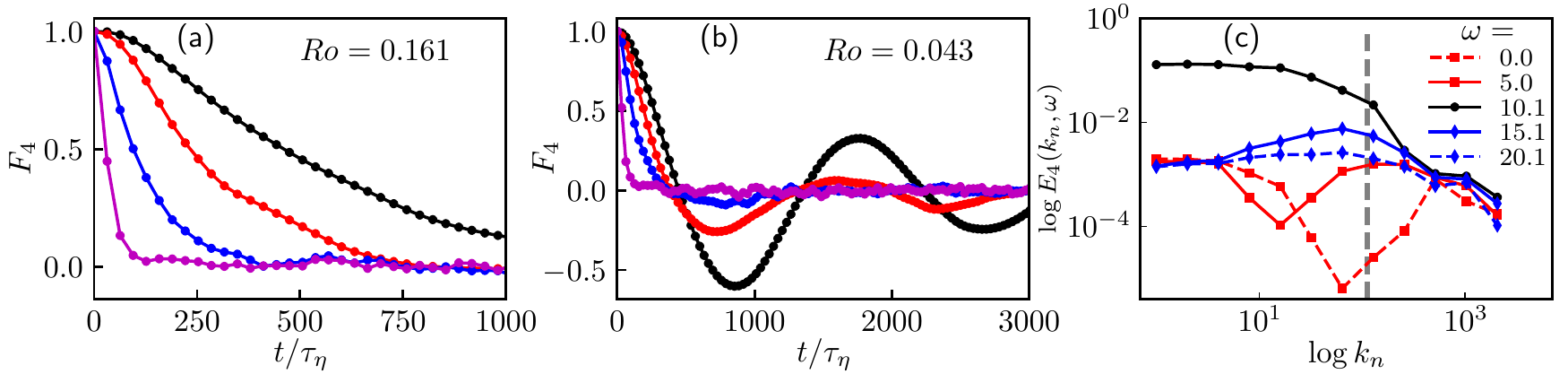}
		\caption{Representative plots for the evolution of normalised time-dependent fourth order structure function $F_4(k_n, t)$ versus time 
		which is normalised by the Kolmogorov time-scale $\tau_\eta$ for (a) $\mathrm{Ro} = 0.161$ and (b) $ \mathrm{Ro} = 0.043$. The black, red, blue and magenta 
		coloured lines correspond to $n = 11$, $n = 12$, $n = 14$ and $n = 16$ respectively. 
		(c) Plots of the relative spectral energy density $E_4(k_n, \omega)$ of $F_4(k_n, t)$ for $ \mathrm{Ro} = 0.043$ versus 
		$k_n$ for different harmonics $\omega$; the vertical dashed line represents the Zeman wavenumber.}
		\label{fig:Fp}
	\end{figure*}	

	\begin{figure}
		\includegraphics[width=1.0\linewidth]{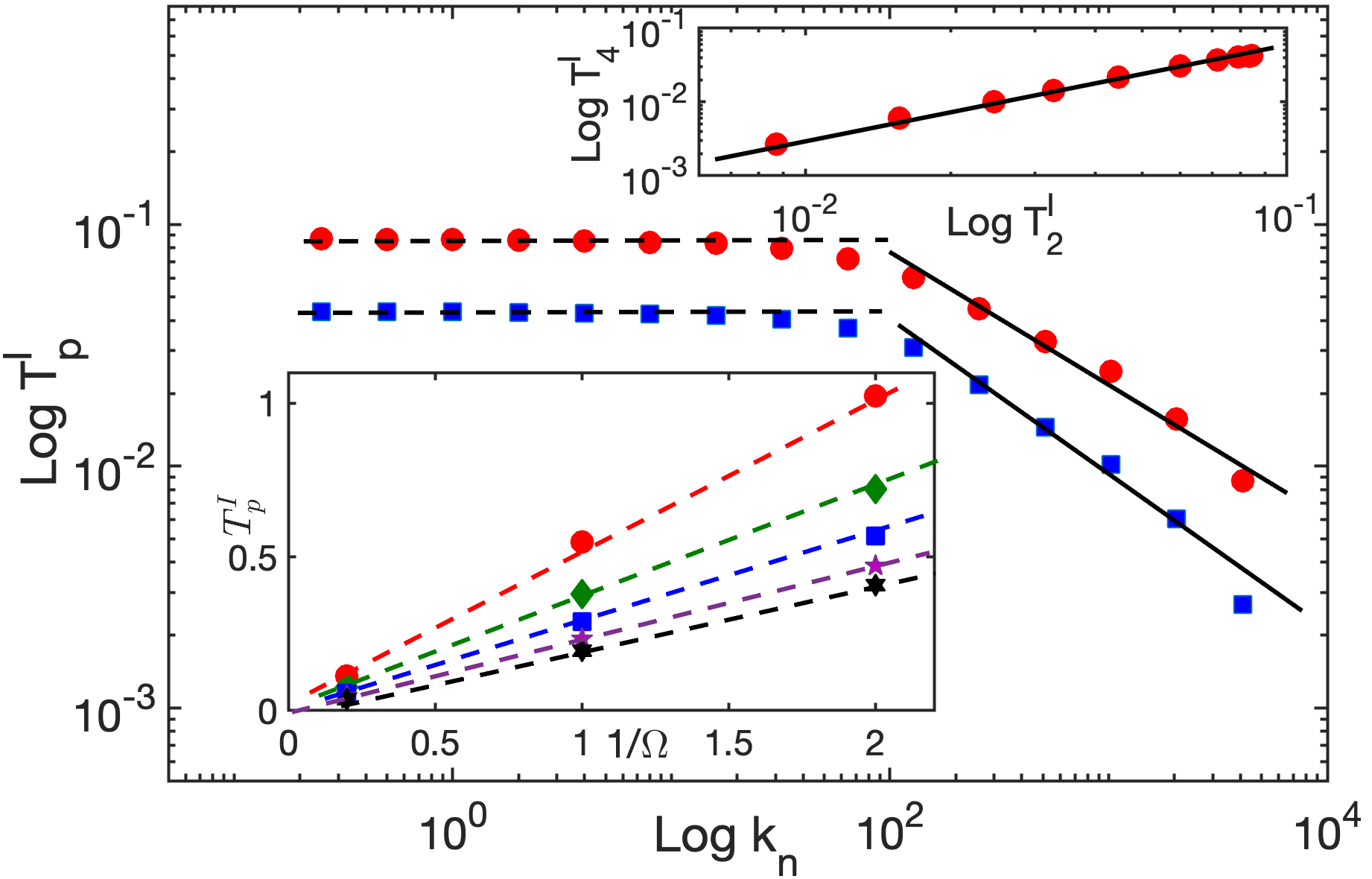}
		\caption{\textcolor{black}{Loglog (base 10) plots of the integral time $T^I_p(k_n)$ versus $k_n$ for $p = 2$ (red circles) and $p = 4$ (blue squares) for $\mathrm{Ro} = 0.043$.
		The dashed horizontal lines are best fits ($z_p = 0$) for wavenumbers lower than the Zeman wavenumber and the thick black lines are the best fits 
		in the inertial range not dominated by rotation ($z_p \neq 0$; see Table 1). Lower Inset: 
		Plots of $T^I_{p}$ ($p =2$ to 6, from the uppermost to the lowermost plot), for $k_n \lesssim k_{\Omega}$, \textit{vs} $1/\Omega$ (except for $\Omega = 0.1$ where the plateau extends for just a few shells) for different values of $p$. The dashed lines are 
		linear fits which show $T^I_{p} \propto 1/\Omega$, consistent with the theoretical prediction.  Upper Inset:  Plot of the fourth vs the second-order integral time-scales shows a convincing scaling with slope $z_4/z_2 \approx 1.09$, consistent 
		with the exponent ratio obtained from the bridge relation.}}
		\label{fig:Tp}
	\end{figure}
	\begin{table*}
		\begin{tabular*}{\linewidth}{c@{\extracolsep{\fill}}  c@{\extracolsep{\fill}}  c@{\extracolsep{\fill}} c@{\extracolsep{\fill}} c@{\extracolsep{\fill}} c@{\extracolsep{\fill}} c@{\extracolsep{\fill}}}
			\hline
			\hline
			 \multirow{2}{*}{$p$}& \multirow{2}{*}{$\zeta_p$} & \multirow{2}{*}{$z_{p}$ (Eq.\ref{exponents})} & \multicolumn{4}{c}{$z_{p}$}\\ \cline{4-7}
			&  &  & $\mathrm{Ro} = 0.809$ & $\mathrm{Ro} = 0.232$ & $\mathrm{Ro} = 0.161$ & $\mathrm{Ro} = 0.043$\\[0.5ex]
			\hline
			1 & 0.379 $\pm$  0.006 & 0.621 $\pm$ 0.006 &0.63 $\pm$ 0.01& 0.64 $\pm$ 0.02& 0.65 $\pm$ 0.03 & 0.67 $\pm$ 0.07\\
			2 & 0.707 $\pm$  0.005 & 0.672 $\pm$ 0.008 & 0.673 $\pm$ 0.009 & 0.68 $\pm$ 0.01   & 0.68 $\pm$ 0.01   &  0.67 $\pm$ 0.06 \\[0.5ex]
			3 & 1.0				   & 0.707 $\pm$ 0.005 & 0.719 $\pm$ 0.009 & 0.703 $\pm$ 0.009 & 0.72 $\pm$ 0.01   & 0.73 $\pm$ 0.04 \\[0.5ex]
			4 & 1.267 $\pm$  0.007 & 0.733 $\pm$ 0.007 & 0.72  $\pm$ 0.01  & 0.72  $\pm$ 0.01  & 0.746 $\pm$ 0.008 & 0.76 $\pm$ 0.02 \\[0.5ex]
			5 & 1.51  $\pm$  0.02  & 0.75  $\pm$ 0.02  & 0.72  $\pm$ 0.02  & 0.74  $\pm$ 0.01  & 0.761 $\pm$ 0.008 & 0.79 $\pm$ 0.02 \\[0.5ex]
			6 & 1.74  $\pm$  0.03  & 0.77  $\pm$ 0.03  & 0.76  $\pm$ 0.02  & 0.75  $\pm$ 0.02  & 0.78 $\pm$ 0.01   & 0.80 $\pm$ 0.03 \\[0.5ex]
			\hline
			\hline
		\end{tabular*}
		\caption{We summarise, for $k_n \gg k_{\Omega}$,  our results for the dynamic exponents $z_p$ (column 3) calculated through the bridge relations~\eqref{exponents} from 
		the equal-time exponents $\zeta_p$ obtained through ESS~\cite{Benzi1993,Chakraborty2010a,Chakraborty2012}(column 2) for different orders $p$ (column 1). Columns 4 - 6 lists the dynamic exponents 
		for different Rossby numbers obtained directly from our shell model simulations. \textcolor{black}{(We note the marginal increase in the error bars and mean 
		exponents, while still being consistent with the theoretical prediction, 
		as $\mathrm{Ro} \to 0$ is likely due to the shrinking of the inertial range $k_n \gg k_{\Omega}$
		as $k_{\Omega}$ becomes larger with decreasing Rossby numbers.)}}
		\label{table:scaling}	
	\end{table*}

Thus, given the question at hand, it is natural for us to approach this problem 
with a shell model for rotating turbulence. Such models are constructed on a logarithmically-spaced 
lattice of wavenumbers $k_n = k_0 \lambda^n$; we use the conventional choices of 
$k_0 = 1/16$ and $\lambda = 2$ in our study. Associated with each shell $n$ is a  
complex variable $u_n$ which mimics velocity increments over a
scale $k_n \sim 1/r$ in the Navier--Stokes equation. By retaining only the nearest 
and next-nearest neighbour couplings in the nonlinear (convolution) term of the Navier--Stokes 
equation, the shell model equations are coupled ordinary differential equations
\begin{eqnarray}\label{goy}
	&&\frac{du_n }{dt}= -\nu k_n^2 u_n + f_n -i\Omega u_n+\nonumber\\&&\,i \left[a k_{n+1} u_{n+2} u_{n+1} + b k_{n} u_{n+1}u_{n-1}  
	+ c k_{n-1} u_{n-1} u_{n-2}\right]^{*}\nonumber\\
\end{eqnarray}
with shell numbers running from 1 to $N$. The asterisk in the equation denotes
a complex conjugation and $i \equiv \sqrt{-1}$ and, as noted before, the
nonlinear couplings are limited ensuring the absence of direct coupling of
large and small scales effectively eliminating sweeping effects. The shell model, in the absence
of viscosity ($\nu = 0$) and external forcing $f_n = 0$, conserves energy,
helicity, and phase-space, through a proper choice of the (real) coefficients
$a$, $b$ and $c$; we use, as is common, $a = 1$, $b = -1/2$, and $c = -1/2$. 

In our simulations of the shell model we choose $N = 27$ shells, $\nu =
10^{-9}$ and an exponential fourth-order Runge-Kutta scheme for time-marching
with a sufficiently small time-step $\delta t = 10^{-5}$ given the stiffness of
these coupled ordinary differential equations.  We use an external forcing on
shells $n = 2$ and $3$ with $f_n = \epsilon (1+i) (\delta_{n,2}/u_n^* +
\delta_{n,3}/2u_n^*)$, where $\epsilon$ specifies the energy input rate and, in our simulations, $\epsilon = 0.01$; 
this specific form  ensures a zero helicity input rate
and the energy flux free from period 3 oscillations~\cite{Kadanoff1995}.  We
initialise our velocity field ($u_n = \sqrt{k_n} \exp(i\theta)$, for $n\le 4$
and $u_n = \sqrt{k_n} \exp(-k_n^2) \exp(i\theta)$  for $n \ge 5$, where $\theta
\in [0,2\pi]$ is a random phase) and drive the system to a statistically steady
state before turning on the Coriolis term.

We characterise rotating, turbulent flows not only by the (large-scale)
Reynolds number $\mathrm{Re} \equiv U_{\rm rms}/k_0\nu$, where the root-mean-square
velocity  $U_{\rm rms}= (\sum_{n} \norm{u_n}^2)^{1/2}$~\cite{Biferale-review},
but also by the Rossby number $\mathrm{Ro} \equiv U_{\rm rms}k_0/\Omega$, which is a
measure of the relative strength of the nonlinearity to the Coriolis force \textcolor{black}{
	and the Zeman wavenumber $k_\Omega = \sqrt{\Omega^3/\varepsilon}$.  We
use, in our simulations, Reynolds number $\mathrm{Re} \sim 10^9$ and 
$\mathrm{Ro} = \infty\, (\Omega = 0;\, k_\Omega = 0), \, 
0.809\, (\Omega = 0.1; \, k_\Omega = 0.3), \, 0.232\, (\Omega = 0.5;\, k_\Omega = 3.5), \, 0.161\, (\Omega = 1.0; \, k_\Omega = 10.0)$, and $0.043\, (\Omega = 5.0;\, k_\Omega = 111.8)$.}

From the statistically steady velocity field of the rotating turbulent flow, it is simple
to define the shell model analogue of the order-$p$, scale-dependent,
quasi-Lagrangian normalised time-dependent structure function as 
\begin{equation}
	F_p(k_n,t) = \Re\frac{\langle [u_n(t_0) u_n^\ast(t_0 + t)]^{p/2} \rangle}{\langle |u_n(t_0)|^{p}\rangle}\label{eq:F_pgoy}
	\end{equation}
where $\Re$ denotes the real part of the function and the angular brackets an average of over different time origins
$t_0$. We choose integer values 
of $p$ between $1$ and $6$ in this study.

In Fig. ~\ref{fig:Fp}(a)  we show representative plots of the fourth-order time-dependent 
structure function $F_4(k_n,t)$, for $\mathrm{Ro} = 0.161$, and different shell numbers which are all greater than  
the Zeman scale and hence much less influenced by the effects of rotation. As one would expect, 
the correlations decay much faster for higher wavenumbers than for lower wavenumbers. 

To bring out the effect of rotation clearly, we go to a lower value of the Rossby number (hence a 
higher value of the Zeman wavenumber). Figure~\ref{fig:Fp}(b) shows such a plot 
for $\mathrm{Ro} = 0.043$ for the same wavenumbers as in panel (a). However, for such a low 
value of $\mathrm{Ro}$, shell numbers $n = 11$ and 12, corresponding to wavenumbers close to the 
Zeman scale, are clearly affected by the Coriolis 
force. This is clearly seen in the conspicuous oscillatory profile of the structure function. 

These oscillations arise, as already shown through the direct numerical
simulations of Eulerian time-dependent correlators in
Refs.~\cite{Favier_2010,Leoni2014}, at rotation-dominated scales $k \lesssim
k_\Omega$ because of the presence of the Coriolis term.  Indeed, the formal
solution of Eq.~(\ref{goy}) ought to have a dominant harmonic
$\sim\exp(-i\Omega t) $, in addition to the contributions of viscosity and the
non-linearity;  this oscillatory factor of course becomes vanishingly small
when $k_n \gg k_\Omega$. Consequently, for $k_n \lesssim k_\Omega$
(Fig.\ref{fig:Fp}(b), $n = 11$ and $n = 12$), the time-dependent structure functions 
$F_p(k_n,t)$ have an oscillatory profile with a dominant harmonic of angular frequency 
$\sim p\Omega t/2$. For wavenumbers much  $k_n \gg k_\Omega$, the nonlinearity of 
the dynamical systems ensures a mixing of the harmonics of different scales results in 
several sub- and super-harmonics in the system eventually washing away the 
clear oscillatory profile seen for $k_n \lesssim k_\Omega$. 

This picture is easily validated, through a Fourier decomposition, from measurements 
of the spectral relative energy content $E_p(k_n, \omega)
\equiv\norm{\hat{F}_p(k_n,\omega)}^2/\sum_{n}\norm{\hat{F}_p(k_n,\omega)}^2$; here 
$\hat{F}_p(k_n,\omega)$ is the Fourier transform of $F_p(k_n, t)$.
Figure~\ref{fig:Fp}(c) illustrates such an analysis, for $p = 4$ and $R = 0.043$ corresponding to the structure functions 
in Fig.\ref{fig:Fp}(b), which clearly shows that while all the energy is maximally contained, for $k_n \lesssim k_\Omega$, 
in the $\omega \sim 10$ (corresponding to $\mathrm{Ro} = 0.043$) mode, at wavenumbers $k_n \gg k_\Omega$ the energy is distributed 
more uniformly amongst the other harmonics that we calculate.

From the time-dependent structure functions of the sort shown in
Fig.~\ref{fig:Fp}(a) and (b), we define the $p$-th order, shell-model analogue
of the integral-time scale $T^I_{p}(k_n) \equiv \int_0^\infty F_p(k_n,t)dt$; in
practice \textcolor{black}{(to avoid contamination from statistical noise at long times~\cite{Mitra2004})}, the upper limit of the integral is restricted to times when
$F_p(k_n,t)$ has reached a value of $0.6$ and we have checked that our results
are insensitive if this limit is varied between $0.4$ and $0.8$. 

\textcolor{black}{In Fig.~\ref{fig:Tp} we show loglog plots of 
$T^I_{2}(k_n)$ and $T^I_{4}(k_n)$ vs $k_n$ for $\mathrm{Ro} = 0.043$. 
Clearly, for $k_n \lesssim
k_\Omega$, the plateau in $T^I_{p}(k_n)$ leads to a dynamic exponent $z_p = 0$ as indicated 
by the dashed best fit lines. In the lower inset, we plot the values of these plateau for different orders vs $1/\Omega$; the 
dashed line fits for each order shows clearly that the theoretical prediction from the multifractal analysis $T^I_p \propto 1/\Omega$, for $k_n \lesssim k_\Omega$ holds.
However for  $k_n \gg k_\Omega$, the
integral-time scale seems to be clearly a power-law with 
$z_p \neq 0$ which extends over a decade as shown by the black lines which best fit the data. 
From plots such as these we extract, through a least-square fit, $z_p$ (for different values
of $\mathrm{Ro}$) from 500 different measurements; in Table 1 we quote the mean of these
exponents and their standard deviations as error bars.} 
\textcolor{black}{To further illustrate the quality of the scaling range for higher wavenumbers, 
in the upper inset we show a loglog plot of the fourth vs the second-order integral time-scale in a manner 
reminiscent of 
the extended self-similarity (ESS)~\cite{Benzi1993,Chakraborty2010a,Chakraborty2012} technique used for equal-time measurements. This representation 
shows a clear scaling with the best fit (black line) slope $z_4/z_2 \approx 1.09$, consistent with what one would obtain from the 
multifractal theory. While this ESS-like approach is convincing, we would advice caution in over-interpreting the 
role of such an extended self-similarity for dynamic exponents in the absence of a theory analogous to what is known for 
equal-time structure functions~\cite{Benzi1993,Chakraborty2010a,Chakraborty2012}.}

Comparing the different columns in Table 1, it is clear that the bridge
relation~\eqref{exponents} are indeed satisfied for all Rossby numbers for
wavenumbers  $k_n \gg k_\Omega$. Furthermore, in the rotation dominated scales
$k_n \lesssim k_\Omega$, we find (within error bars) $z_p = 0$, again
consistent with our theoretical prediction~\eqref{exponents}. 

In this paper we have addressed the issue of dynamic scaling in rotating turbulence by using the tools 
of the Frisch-Parisi multifractal formalism and then validated our predictions through detailed numerical 
simulations of a shell model which factors in the Coriolis force. By adopting a quasi-Lagrangian approach, 
our work complements earlier (Eulerian) studies~\cite{Favier_2010,Leoni2014} of time-dependent correlation functions in such flows. 
We uncover a new set of exponents~\eqref{exponents}, 
and associated bridge relations and find, unsurprisingly, for wavenumbers larger than the Zeman scale, even strongly rotating 
flows show dynamic multiscaling which is completely consistent to what has been known~\cite{Mitra2004,Ray2008}. Surprisingly, 
for wavenumbers which are dominated by the rotation, the relevant time-scales are scale-independent and thence $z_p = 0$ \textcolor{black}{in sharp 
contrast to estimates from local time-scale arguments.
This is because at such scales rotation is the dominant mechanism when compared 
with those imposed by the nonlinear term. Hence na\"ively one would expect that the dominant time-scale here 
would be $\sim 1/\Omega$; the multifractal approach picks out this dominant time-scale over all others. 
This is perhaps because at 
these scales a local turnover time approach fails to factor in
the time-scale imposed on the flow by rotation which dominates over the intrinsic (and local) 
time-scales arising in the flow itself.
Thus a comparison of the time-scales which emerge from arguments based on the  turnover 
time with those from the multifractal model shows a greater disparity at rotation dominated scales than those 
which are not underlining the singular nature of the Coriolis force especially when it comes to dynamic correlators.}
Furthermore, curiously the intermittency corrections seen in the equal-time 
measurements seem to be absent from the dynamics altogether. To the best of our knowledge this is the only example of a turbulent flow 
where such a decoupling of a fundamental feature of turbulence happens when we move from the statics to the dynamics and deserves 
a more rigorous investigation in the future.

\begin{acknowledgements} SSR acknowledges the support of SERB-DST (India) projects MTR/2019/001553,  STR/2021/000023, CRG/2021/002766 and the DAE, Govt. of India, under project no.  12-R\&D-TFR-5.10-1100 and project no.  RTI4001.
\end{acknowledgements}

\bibliography{rathor_etal_manuscript}
\end{document}